\documentclass[review]{elsarticle}

\usepackage{lineno}
\usepackage[colorlinks=true,linkcolor=black,citecolor=blue,urlcolor=blue,pdfauthor=author]{hyperref}
\modulolinenumbers[5]
\usepackage{amsmath,amsfonts,amssymb,amsthm,bm}
\usepackage{float}
\usepackage{graphics}
\usepackage{graphicx}
\usepackage{caption}
\graphicspath{{figures/}}
\usepackage{subfigure}
\usepackage{fullpage}
\usepackage{multirow}
\usepackage{booktabs}
\usepackage{textcomp}
\usepackage{gensymb}
\usepackage{nomencl}
\usepackage{comment}
\usepackage{setspace}
\usepackage{geometry}
\usepackage[section]{placeins}
\usepackage{soul}

\makeatletter
\def\ps@pprintTitle{%
	\let\@oddhead\@empty
	\let\@evenhead\@empty
	\def\@oddfoot{}%
	\let\@evenfoot\@oddfoot}
\makeatother

\begin{document}
\captionsetup[figure]{labelfont={bf},name={Fig.},labelsep=period}
\begin{frontmatter}

\title{A multiple scattering formulation for finite-size flexural metasurfaces}

\author[mymainaddress] {Xingbo Pu}
\author[mymainaddress] {Antonio Palermo\corref{mycorrespondingauthor}}

\author[mymainaddress] {Alessandro Marzani\corref{mycorrespondingauthor}}
\cortext[mycorrespondingauthor]{Corresponding author}

\ead{antonio.palermo6@unibo.it; alessandro.marzani@unibo.it}

\address[mymainaddress]{Department of Civil, Chemical, Environmental and Materials Engineering, University of Bologna, 40136 Bologna, Italy}

\begin{abstract}

We provide an analytical formulation to model the propagation of elastic waves in a homogeneous half-space supporting an array of thin plates. The
technique provides the displacement field obtained from the interaction between an incident wave generated by a harmonic source and the scattered fields induced by the flexural motion of the plates. The scattered field generated by each plate is calculated using an ad-hoc set of Green's functions. The interaction between the incident field and the scattered fields is modeled through a multiple scattering formulation. Owing to the introduction of the multiple scattering formalism, the proposed technique can handle a generic set of plates arbitrarily arranged on the half-space surface. The method is validated via comparison with finite element simulations considering Rayleigh waves interacting with a single and a collection of thin plates. Our framework can be used to investigate the interaction of vertically polarized surface waves and flexural resonators in different engineering contexts, from the design of novel surface acoustic wave devices to the interpretation of urban vibrations problems. 

\end{abstract}


\end{frontmatter}


\section{Introduction}

In the past two decades, the advent of so-called metamaterials, structured materials equipped with resonant elements, has opened new horizons in the control of waves across different physics. In elastodynamics, architected materials with unique effective properties such as negative density and negative moduli have shown the capability of tailoring the wavefields at will \cite{hussein2014dynamics}.

At their conception, elastic metamaterials were designed to filter the propagation of bulk waves \cite{liu2000locally}. The fabrication of resonant elements in the host medium, however, challenges the feasibility of large-scale samples of such materials and eventually disrupts the integrity of the hosting material to some extent \cite{lim2021photonic}.

These issues were partially mitigated by the design of elastic metasurfaces, thin resonant interfaces comprising subwavelength resonant elements located on the surface of the waveguides \cite{boechler2013interaction,colombi2016forests,colquitt2017seismic,palermo2020surface}. Nowadays, metasurfaces are proposed to manipulate waves across a wide range of scales, from micro-mechanical systems \cite{raguin2019dipole} to seismic contexts \cite{colquitt2017seismic,pu2020seismic, zeighami2021rayleigh}. Understanding the dynamics of a metasurface requires, at least, the definition of its dispersion relation. Knowledge of the dispersion curves provides physical insights on the existence of bandgaps, modes hybridization phenomena, cut-on and cut-off frequencies of an infinite length system. Dispersion relations of metasurfaces can be computed either via analytical formulations, by modeling the thin resonant interface as dynamic boundary conditions (BCs)  \cite{garova1999interaction, boechler2013interaction}, or via finite element (FE) schemes by modeling a finite size portion of the system including the substrate and the metasurface and by imposing proper periodic BCs \cite{khelif2010locally, veres2012complexity, palermo2018metabarriers}. The analytical formulations often exploit homogenization and asymptotic expansion techniques to obtain closed-form dispersion relation
for compressional (rod-like) \cite{colquitt2017seismic, wootton2020second} and flexural (beam-like) metasurfaces
\cite{wootton2019asymptotic, marigo2020surface}. 

Transmission coefficients and full wavefields provide the additional information required to characterize the dynamics of a finite length metasurface and evidence complex wave patterns like lensing \cite{palermo2018control, fuentes2021design}, classical \cite{maurel2018conversion} or umklapp \cite{chaplain2020tailored} mode conversion, rainbow trapping \cite{colombi2016seismic} and wave localization \cite{ungureanu2021localizing}. Transmission coefficients and wavefields are generally computed out of numerical harmonic or time transient simulations, by using tools like finite elements. Given the nature of the wave problems, such simulations are always bounded by their computational cost and often limited by the available hardware resources. Analytical treatments are currently limited to ideal configurations, where identical resonators are arranged in a periodic manner (infinite waveguide) so that the transmission coefficients can be obtained in closed forms exploiting homogenization techniques \cite{boutin2006wave, schwan2016site, marigo2020effective}.

Hence, analytical strategies to model metasurfaces of finite-size composed by non-identical resonators are desirable. In a recent work \cite{PU2021103547}, we proposed a multiple scattering formulation to model Rayleigh waves interacting with a finite length metasurface placed atop an elastic half-space. The considered metasurface comprised an arbitrary number of discrete mass-spring resonators, oscillating along their vertical direction (compressional resonance). In this work we extend the previous formulation by considering flexural-type resonators, namely thin Kirchhoff plates, coupled to the half-space. The formulation is able to account for an arbitrary number $N$ of thin plates. We focus on the low-frequency regime where the flexural contribution of the plates dominates the dynamic response. Thus, we neglect the plates axial behavior, and the related axial resonances which generally occur at a much higher frequency w.r.t the first bending modes.

The key point of our formulation is the solution of the Lamb's problem in terms of Green's functions. Such Green's functions are used to model the incident wavefield, due to a known harmonic source acting normally at the surface of the half-space, as well as the scattered wavefields generated by the thin plates when excited by an imposed base displacement. The unknown amplitudes of the $N$ scattered wavefields are found from the solution of the proposed multiple scatting formulation by imposing proper boundary conditions at the half-space surface.
The total wavefield in the half-space is thus obtained by the coupled contribution of the incident and $N$ scattered wavefields.

Our approach enriches the analytical tools available to discuss the dynamics of flexural metasurfaces, e.g., the closed form dispersion relationships in \cite{wootton2019asymptotic} and the effective models in \cite{schwan2016site,marigo2020effective}, by enabling the description of finite-size flexural metasurface. To the best of our knowledge, the analytical treatment of this problem remains yet unsolved.

The work is organized as follows. In Section \ref{Theoretical model}, we present the derivation of the theoretical framework. In particular, in Section \ref{Impedance boundary conditions of thin plates on an elastic half-space} we begin by recalling the Kirchhoff plate theory to derive the shear and normal stresses at the base of the plates as impedance functions. Then, we recall and use the solution of the Lamb's problem to formulate the incident and scattered wavefields in Section \ref{Green's functions}. The multiple scattering formulation is finally constructed in Section \ref{Multiple scattering formulation}. In Section \ref{Numerical examples}, we investigate and discuss the response of an half-space equipped with a single and a collection of thin plates, and validate our findings via FE simulations. We conclude our work with some remarks and perspectives in Section \ref{Conclusion}.

\section{Theoretical model} \label{Theoretical model}

\begin{figure}[htbp]
	\centering
	\includegraphics[width=5 in]{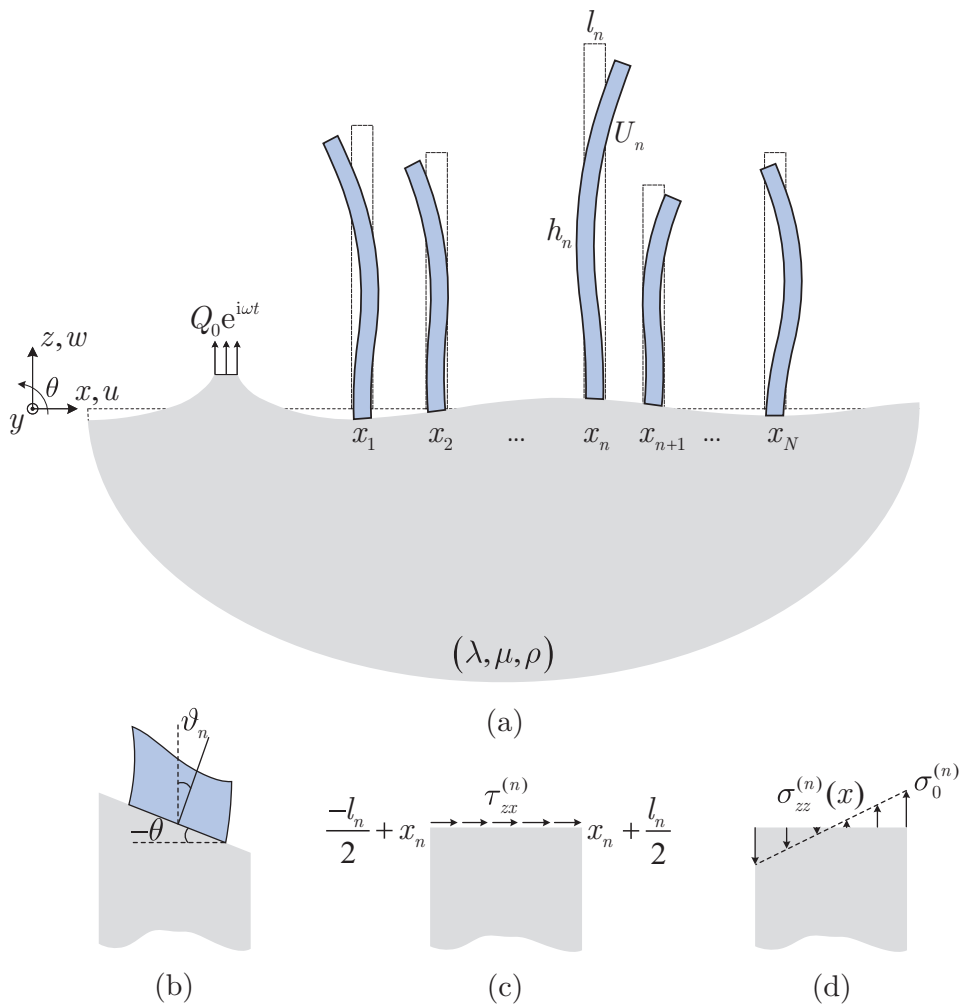}
	\caption{Schematic of elastic waves interacting with a metasurface. (a) An array of arbitrary $N$ thin plates on an elastic half-space, in which $h_n$ and $l_n$ denote the height and thickness of the $n$-th plate, respectively. (b) Illustration of the continuity of displacements and slope at the plate-substrate interface. The assumed shear and normal stresses caused by the flexural motion of the $n$-th plate are shown in (c) and (d), respectively.}
	\label{fig:fig1}
\end{figure}

We investigate the multiple scattering effect of an array of flexural resonators, aka a flexural metasurface, arranged on the surface of an elastic half-space, as shown in Figure \ref{fig:fig1}. The flexural metasurface can be composed by a series of $N$ parallel thin plates, with different mechanical and geometrical properties. We model the plates using the Kirchhoff plate theory and their effect on the substrate dynamics using ad-hoc impedance functions. We consider a two-dimensional (2D) model, given the invariance of the problem in the out-of plane direction, in the frequency domain so that the time-harmonic term $\mathrm{e}^{\mathrm{i} \omega t}$ is omitted through the whole derivation.

The derivation comprises the following three steps: (i) definition of the impedance functions describing the stresses exerted by a thin plate at its base when excited by an imposed base motion; (ii) construction of the half-space Green's functions for such stresses; (iii) formulation of a multiple scattering problem for a half-space coupled with an arbitrary $N$ number of thin plates under a harmonic strip source at the surface. 

In the 2D plane described by the  spatial coordinates $\boldsymbol{x}=(x,z)$ , we denote the displacement components in the half-space by $u(x,z)$ and $w(x,z)$ along the $x$ and $z$ directions, respectively. The displacement component of the thin plate along the $x$ direction is $U(x,z)$ whereas $\vartheta(x,z)=\partial U(x,z)/\partial z$ denotes the cross-section rotation. For convenience, we introduce a set $\mathcal{B}_n=\{\boldsymbol{x}: x=x_n, 0\le z \le h_n, n \in \mathbb{Z}^+\} \subset \mathbb{R}^2$ to indicate the $n$-th plate placed at $x_n$, such that its displacements components read $U_n(z)$ and $\vartheta_n(z)$, for $n=1,..,N$.  

\subsection{Impedance boundary conditions of thin plates on an elastic half-space} \label{Impedance boundary conditions of thin plates on an elastic half-space}

The thin plates are made of homogeneous and linear elastic materials, and are perfectly bonded to the surface of an elastic half-space. The flexural motion of the $n$-th plate is governed by the equation:
\begin{equation} \label{equ:bending motion}
\left(\frac{\partial^4}{\partial z^4}-\beta_n^4\right)U_n=0, \quad \text{for} \quad \boldsymbol{x} \in \mathcal{B}_n
\end{equation}
\noindent where $\beta_n$ denotes the wave number for flexural waves, namely:
\begin{equation} 
\beta_n^4=\frac{\rho_n l_n}{D_n}\omega^2, \quad \text{where} \quad D_n = \frac{E_n}{1-\nu_{n}^2}\frac{l_{n}^3}{12}
\end{equation}
\noindent in which $\rho_n$ is the material density, $l_n$ is the plate thickness, $D_n$ is the flexural rigidity, $E_n$ is the Young modulus, and $\nu_n$ is the Poisson ratio. The general solution of (\ref{equ:bending motion}) has the following form:

\begin{equation} \label{equ:general solution of flexural motion}
U_n(z)=C_{n1} \cosh (\beta_n z)+C_{n2} \sinh (\beta_n z)+C_{n3} \cos (\beta_n z)+C_{n4} \sin (\beta_n z).
\end{equation}
\noindent in which the four coefficients $C_{ni}\; (i=1,2,3,4)$ can be determined by imposing the four boundary conditions (BCs) for flexural vibrations:

\begin{subequations} 
	\begin{equation} \label{equ:BC of flexural motion at z=0}
	U_n(0)=u(x_n,0), \quad  \vartheta_n(0)=-\theta(x_n,0).
	\end{equation}
	\begin{equation} \label{equ:BC of flexural motion at z=h}
	\frac{\partial^2 U_n}{\partial z^2}\big|_{z=h_n}=0, \quad \frac{\partial^3 U_n}{\partial z^3}\big|_{z=h_n}=0. 
	\end{equation}
\end{subequations}
where the slope of the free surface $\theta(x_n,0)$ can be expressed in terms of the vertical displacement in the soil $w$ as $\frac{\partial w(x,0)}{\partial x}\big|_{x=x_n}$.

Equation (\ref{equ:BC of flexural motion at z=0}) imposes the continuity of displacement and slope at the interface between the $n$-th plate and the half-space, while (\ref{equ:BC of flexural motion at z=h}) implies null bending moment and shear force at the plate free end. Expressing (\ref{equ:BC of flexural motion at z=0}) and (\ref{equ:BC of flexural motion at z=h}) in terms of (\ref{equ:general solution of flexural motion}) leads to the following non-homogeneous system:

\begin{equation} \label{equ:matrix of flexural motion BCs}
\left[\begin{array}{cccc}
1 & 0 & 1 & 0 \\
0 & 1 & 0 & 1 \\
\cosh (\beta_n h_n) & \sinh (\beta_n h_n) & -\cos (\beta_n h_n) & -\sin (\beta_n h_n) \\
\sinh (\beta_n h_n) & \cosh (\beta_n h_n) & \sin (\beta_n h_n) & -\cos (\beta_n h_n)
\end{array}\right]\left[\begin{array}{c}
C_{n1} \\
C_{n2} \\
C_{n3} \\
C_{n4}
\end{array}\right]=\left[\begin{array}{c}
u(x_n,0) \\
-\theta(x_n,0) / \beta_n \\
0 \\
0
\end{array}\right]
\end{equation}
which can be solved for the four unknowns $C_{ni}$:

\begin{subequations}
	\begin{equation} \label{equ:coefficient: C_1}
	\begin{split}
	C_{n1}=& \frac{1+\cosh (\beta_n h_n) \cos (\beta_n h_n)+\sinh (\beta_n h_n) \sin (\beta_n h_n)}{2+2 \cosh (\beta_n h_n) \cos (\beta_n h_n)} u(x_n,0)\\
	&- \frac{\cosh (\beta_n h_n) \sin (\beta_n h_n)-\sinh (\beta_n h_n) \cos (\beta_n h_n)}{2+2 \cosh (\beta_n h_n) \cos (\beta_n h_n)} \frac{\theta(x_n,0)}{\beta_n}, 
	\end{split}
	\end{equation}
	
	\begin{equation} \label{equ:coefficient: C_2}
	\begin{split}
	C_{n2}=& -\frac{\cosh (\beta_n h_n) \sin (\beta_n h_n)+\sinh (\beta_n h_n) \cos (\beta_n h_n)}{2+2 \cosh (\beta_n h_n) \cos (\beta_n h_n)} u(x_n,0)\\
	&- \frac{1+\cosh (\beta_n h_n) \cos (\beta_n h_n)-\sinh (\beta_n h_n) \sin (\beta_n h_n)}{2+2 \cosh (\beta_n h_n) \cos (\beta_n h_n)} \frac{\theta(x_n,0)}{\beta_n},
	\end{split}
	\end{equation}
	
	\begin{equation} \label{equ:coefficient: C_3}
	\begin{split}
	C_{n3}=& \frac{1+\cosh (\beta_n h_n) \cos (\beta_n h_n)-\sinh (\beta_n h_n) \sin (\beta_n h_n)}{2+2 \cosh (\beta_n h_n) \cos (\beta_n h_n)} u(x_n,0) \\
	&- \frac{\sinh (\beta_n h_n) \cos (\beta_n h_n)-\cosh (\beta_n h_n) \sin (\beta_n h_n)}{2+2 \cosh (\beta_n h_n) \cos (\beta_n h_n)} \frac{\theta(x_n,0)}{\beta_n},
	\end{split}
	\end{equation}
	
	\begin{equation} \label{equ:coefficient: C_4}
	\begin{split}
	C_{n4} &=\frac{\cosh (\beta_n h_n) \sin (\beta_n h_n)+\sinh (\beta_n h_n) \cos (\beta_n h_n)}{2+2 \cosh (\beta_n h_n) \cos (\beta_n h_n)} u(x_n,0) \\
	&- \frac{1+\cosh (\beta_n h_n) \cos (\beta_n h_n)+\sinh (\beta_n h_n) \sin (\beta_n h_n)}{2+2 \cosh (\beta_n h_n) \cos (\beta_n h_n)} \frac{\theta(x_n,0)}{\beta_n}.
	\end{split}
	\end{equation}
\end{subequations}

Given the displacement $U_n(z)$, the shear and normal stresses on the contact area of the $n$-th plate, i.e. in the region $x\in (x_n-l_n/2, x_n+l_n/2)$ and $z=0$, can be approximated as (see Figures \ref{fig:fig1}c and \ref{fig:fig1}d):

\begin{subequations}
	\begin{equation} \label{equ:shear stress from bending moment}
	\begin{split}
	\tau_{zx}^{(n)}(x,0)&=-\frac{D_n}{l_n} \frac{\partial^3 U_n}{\partial z^3} = \frac{D_n}{l_n} \beta_n^3(C_4-C_2) \\
	&=\Omega_1^{(n)} u(x_n,0)-\Omega_2^{(n)} \theta(x_n,0),
	\end{split}
	\end{equation}
	
	\begin{equation} \label{equ:normal stress from bending moment}
	\begin{split}
	\sigma_{zz}^{(n)}(x,0)&=-\frac{E_n}{(1-\nu_{n}^2)} \frac{\partial^2 U_n}{\partial z^2}(x-x_n) = \frac{E_n}{(1-\nu_{n}^2)} \beta_n^2(C_3-C_1)(x-x_n) \\
	&=[\Omega_3^{(n)} u(x_n,0)-\Omega_4^{(n)} \theta(x_n,0)](x-x_n).
	\end{split}
	\end{equation}
\end{subequations}
\noindent
where the following parameters have been defined:
\begin{subequations}
	\begin{equation} \label{equ:Omega_1}
	\Omega_1^{(n)} = \frac{D_n}{l_n} \beta_n^3 f_1(\beta_n h_n),
	\end{equation}
	\begin{equation} \label{equ:Omega_2}
	\Omega_2^{(n)} = \frac{D_n}{l_n} \beta_n^2 f_2(\beta_n h_n),
	\end{equation}
	\begin{equation} \label{equ:Omega_3}
	\Omega_3^{(n)} = -\frac{E_n}{1-\nu_{n}^2} \beta_n^2 f_3(\beta_n h_n),
	\end{equation}
	\begin{equation} \label{equ:Omega_4}
	\Omega_4^{(n)} = -\frac{E_n}{1-\nu_{n}^2} \beta_n f_4(\beta_n h_n).
	\end{equation}
\end{subequations}
\noindent
with:
\begin{subequations}
	\begin{equation}
	f_1(\beta_n h_n) = \frac{\cosh (\beta_n h_n) \sin (\beta_n h_n)+\sinh (\beta_n h_n) \cos (\beta_n h_n)}{1+\cosh (\beta_n h_n) \cos (\beta_n h_n)}, 
	\end{equation}
	\begin{equation}
	f_2(\beta_n h_n) = f_3(\beta_n h_n)=\frac{\sinh (\beta_n h_n) \sin (\beta_n h_n)}{1+\cosh (\beta_n h_n) \cos (\beta_n h_n)}, 
	\end{equation}
	\begin{equation}
	f_4(\beta_n h_n) = \frac{\cosh (\beta_n h_n) \sin (\beta_n h_n)-\sinh (\beta_n h_n) \cos (\beta_n h_n)}{1+\cosh (\beta_n h_n) \cos (\beta_n h_n)}. 
	\end{equation}
\end{subequations}

In short, equations (\ref{equ:shear stress from bending moment}) and (\ref{equ:normal stress from bending moment}) provide the shear and normal stresses at the base of the $n$-th plate for a given imposed harmonic motion at its base.

Since the system considers an arbitrary number $N$ of plates, the total wavefield must be determined considering the incident wavefield actuated by the given source plus the contributions of the $N$ scattered wavefields generated by the plates. In what follows, we set a multiple scattering problem, exploiting the Green's function of the Lamb's problem for the incident and scattered waves, and solve it to find the unknown amplitudes of the $N$ scattered wavefields. We show that such amplitudes can be determined by imposing continuity conditions between the impedance functions of the plates, derived above, and the half-space.

\subsection{Green's functions } \label{Green's functions}
Let us briefly recall the fundamental steps of the Lamb's problem \cite{lamb1904propagation} to provide the Green's functions in terms of displacement components ($u$ and $w$), for given time-harmonic shear and normal stress distributions, $\tau$ and $\sigma$ respectively, acting at the top surface ($z=0$) of the half-space.

At first, we consider the  potentials $\varphi(x,z)$, $\psi_y(x,z)$ which satisfy the wave equations \cite{achenbach1973wave}:
\begin{equation} \label{equ:wave equations}
(\nabla^2 + k_p^2)\varphi=0,  \quad  (\nabla^2 + k_s^2)\psi_y=0.
\end{equation}
\noindent
in which $k_p=\frac{\omega}{c_p}$ and $k_s=\frac{\omega}{c_s}$ denote the wave numbers for compression and shear waves in the half-space, respectively, and where:
\begin{equation} \label{equ:wave velocities}
c_p=\sqrt{\frac{\lambda+2 \mu}{\rho}}, \quad c_s=\sqrt{\frac{\mu}{\rho}}.
\end{equation}
are the compression and shear wave velocities, respectively, being $\lambda$ and $\mu$ the Lam\'e constants and $\rho$ the mass density of the half-space.

Making use of the spatial Fourier transform pairs:

\begin{equation} \label{equ:definition of FT}
\bar{f}(k,z)=\int_{-\infty}^{\infty}f(x,z) \mathrm{e}^{-\mathrm{i}kx}\,\mathrm{d}x,\quad      f(x,z)=\frac{1}{2\pi} \int_{-\infty}^{\infty}\bar{f}(k,z) \mathrm{e}^{\mathrm{i}kx}\,\mathrm{d}k.
\end{equation}
\noindent
we obtain the transformed wave equations (\ref{equ:wave equations}) as:

\begin{equation}
\left(\frac{\partial^2}{\partial z^2}-p^2\right) \bar{\varphi}=0, \quad\left(\frac{\partial^2}{\partial z^2}-q^2\right) \bar{\psi}_y=0.
\end{equation}
which admit solutions of the form:

\begin{equation} \label{equ:transformed potentials}
\bar{\varphi}(k,z)=B_1 \mathrm{e}^{pz}, \quad \bar{\psi}_y(k,z)=B_2 \mathrm{e}^{qz}.
\end{equation}
where $p$ and $q$ read:
\begin{equation} \label{equ:wave numbers of potentials}
p=\sqrt{k^2-k_p^2}, \quad q=\sqrt{k^2-k_s^2}.
\end{equation}
\noindent and where the coefficients $B_1, B_2$ are determined by imposing stress BCs at $z=0$. 

To this purpose, we first express the normal and shear stresses in the half-space in terms of the potentials:
\begin{subequations}
	\begin{equation} \label{equ:normal stress in the half-space}
	\sigma_{zz}(x,z)=-\mu \left[k_s^{2} \varphi + 2\left(\frac{\partial^2 \varphi}{\partial x^2} -\frac{\partial^2 \psi_y}{\partial x \partial z}\right)\right], 
	\end{equation}
	\begin{equation} \label{equ:shear stress in the half-space}
	\tau_{zx}(x,z)=-\mu \left[k_s^{2} \psi_y - 2\left(\frac{\partial^2 \varphi}{\partial x \partial z} -\frac{\partial^2 \psi_y}{\partial z^2}\right)\right].
	\end{equation}
\end{subequations}

Next, we Fourier transform the stress components in (\ref{equ:normal stress in the half-space}) and (\ref{equ:shear stress in the half-space}) and impose their values at $z=0$ equal to those given by the source function as:

\begin{subequations}
	\begin{equation} \label{equ:normal stress FT}
	\bar{\sigma}_{zz}(k,0)=\mu[(2k^2-k_s^2) B_1 + 2\mathrm{i}kq B_2]=\bar{\sigma}_{zz}^{(s)}(k,0),
	\end{equation}
	\begin{equation} \label{equ:shear stress FT}
	\bar{\tau}_{zx}(k,0)=\mu[2\mathrm{i}kp B_1 - (2k^2-k_s^2)] B_2 =\bar{\tau}_{zx}^{(s)}(k,0),
	\end{equation}
\end{subequations}
where the superscript ($s$) denotes the stresses generated by the source function. Solving (\ref{equ:normal stress FT}) and (\ref{equ:shear stress FT}) yields the coefficients:
\begin{subequations}
	\begin{equation} \label{equ:potential coefficient B1}
	B_1 =\frac{(2\mathrm{i}kq) \bar{\tau}_{zx}^{(s)}(k,0)+(2k^2-k_s^2)\bar{\sigma}_{zz}^{(s)}(k,0)}{\mu R(k)},
	\end{equation}
	\begin{equation} \label{equ:potential coefficient B2}
	B_2 =\frac{-(2k^2-k_s^2)\bar{\tau}_{zx}^{(s)}(k,0)+(2\mathrm{i}kp) \bar{\sigma}_{zz}^{(s)}(k,0)}{\mu R(k)},
	\end{equation}
\end{subequations}
where $R(k)$ is known as the Rayleigh function:
\begin{equation} 
R(k)=(2k^2-k_s^2)^2-4k^2 pq. 
\end{equation}

According to the Helmholtz decomposition, the displacement components in the half-space can be expressed as:
\begin{equation} \label{equ:displacement expression in the half-space}    
u=\frac{\partial \varphi}{\partial x}-\frac{\partial \psi_{y}}{\partial z}, \quad  w=\frac{\partial \varphi}{\partial z}+\frac{\partial \psi_{y}}{\partial x}.
\end{equation}

Fourier transforming (\ref{equ:displacement expression in the half-space}), and substituting (\ref{equ:transformed potentials}), (\ref{equ:potential coefficient B1}) and (\ref{equ:potential coefficient B2}), yield:

\begin{subequations}
	\begin{equation} \label{equ:transformed u}
	\begin{split}
	\bar{u}(k,z) &= \mathrm{i}k \bar{\varphi}-\frac{\partial \bar{\psi}_y}{\partial z} \\
	&= \frac{-2k^2q\mathrm{e}^{pz}+q(2k^2-k_s^2)\mathrm{e}^{qz}}{\mu R(k)}\bar{\tau}_{zx}^{(s)}(k,0)+\frac{\mathrm{i}k(2k^2-k_s^2)\mathrm{e}^{pz}-(2\mathrm{i}kpq)\mathrm{e}^{qz}}{\mu R(k)}\bar{\sigma}_{zz}^{(s)}(k,0),
	\end{split}
	\end{equation}
	
	\begin{equation} \label{equ:transformed w}
	\begin{split}
	\bar{w}(k,z) &= \frac{\partial \bar{\varphi}}{\partial z} + \mathrm{i}k \bar{\psi}_y \\
	&= \frac{(2\mathrm{i}kpq)\mathrm{e}^{pz}-\mathrm{i}k(2k^2-k_s^2)\mathrm{e}^{qz}}{\mu R(k)}\bar{\tau}_{zx}^{(s)}(k,0)+\frac{p(2k^2-k_s^2)\mathrm{e}^{pz}-2k^2p\mathrm{e}^{qz}}{\mu R(k)}\bar{\sigma}_{zz}^{(s)}(k,0).
	\end{split}
	\end{equation}
\end{subequations}

At last, the inverse Fourier transform of (\ref{equ:transformed u}) and (\ref{equ:transformed w}) provides the wavefield displacement components due to time-harmonic stresses imposed at the surface as:

\begin{subequations}
	\begin{equation} \label{equ:general solution of u}
	u(x,z)=\frac{1}{2\pi \mu} \int_{-\infty}^{\infty}\left[  \frac{-2k^2q\mathrm{e}^{pz}+q(2k^2-k_s^2)\mathrm{e}^{qz}}{R(k)}\bar{\tau}_{zx}^{(s)}(k,0)+\frac{\mathrm{i}k(2k^2-k_s^2)\mathrm{e}^{pz}-(2\mathrm{i}kpq)\mathrm{e}^{qz}}{R(k)}\bar{\sigma}_{zz}^{(s)}(k,0)\right] \mathrm{e}^{\mathrm{i}kx}\,\mathrm{d}k, 
	\end{equation}
	\begin{equation} \label{equ:general solution of w}
	w(x,z)=\frac{1}{2\pi \mu} \int_{-\infty}^{\infty}\left[  \frac{(2\mathrm{i}kpq)\mathrm{e}^{pz}-\mathrm{i}k(2k^2-k_s^2)\mathrm{e}^{qz}}{R(k)}\bar{\tau}_{zx}^{(s)}(k,0)+\frac{p(2k^2-k_s^2)\mathrm{e}^{pz}-2k^2p\mathrm{e}^{qz}}{R(k)}\bar{\sigma}_{zz}^{(s)}(k,0)\right] \mathrm{e}^{\mathrm{i}kx}\,\mathrm{d}k.
	\end{equation}
\end{subequations}

In what follows, we specialize these Green's functions for the stress distributions considered in this work, namely (i) a uniform normal stress distribution, used to model the external source (see Figure \ref{fig:fig1}a), and (ii) a constant shear stress distribution and (iii) a butterfly normal stress distribution, as shown in Figure \ref{fig:fig1}c and Figure \ref{fig:fig1}d, respectively, used to model the stress components generated by the flexural motion of the plates. 

\subsubsection{Uniform normal stress distribution to model the source}

We consider a vertical distributed source with footprint width $l_s$ centered at the origin of the reference system, as shown in Figure \ref{fig:fig1}a. The BCs on the free surface ($z=0$) are:

\begin{equation} \label{equ:uniform normal stress BC at z=0}    
\sigma_{zz} (x, 0) = \left\{
\begin{array}{rl}
Q_0 & \text{if}\, |x| \le l_s/2\\
0 & \text{elsewhere}
\end{array}, \right. \quad
\tau_{zx} (x, 0) = 0,
\end{equation}
\noindent where $Q_0$ is the magnitude of the source. Fourier transforming (\ref{equ:uniform normal stress BC at z=0}) yields:

\begin{equation} \label{equ:FT uniform normal stress BC at z=0}    
\bar{\sigma}_{zz} (k, 0) = \frac{2Q_0}{k}\sin(kl_s/2), \quad
\bar{\tau}_{zx} (k, 0) = 0.
\end{equation}
Substituting (\ref{equ:FT uniform normal stress BC at z=0}) into (\ref{equ:general solution of u}) and (\ref{equ:general solution of w}) we can obtain the displacement components due to a unitary normal stress ($Q_0=1$ Pa) acting at the free surface, namely the Green's functions:

\begin{subequations}
	\begin{equation} \label{equ:G_zu}
	G_{\sigma u}(x,z)=\frac{\mathrm{i}}{\pi \mu} \int_{-\infty}^{\infty}\sin(kl_s/2) \frac{(2k^2-k_s^2)\mathrm{e}^{pz}-2pq\mathrm{e}^{qz}}{R(k)} \mathrm{e}^{\mathrm{i}kx}\,\mathrm{d}k, 
	\end{equation}
	\begin{equation} \label{equ:G_zw}
	G_{\sigma w}(x,z)=\frac{1}{\pi \mu} \int_{-\infty}^{\infty}\frac{\sin(kl_s/2)}{k}\frac{p(2k^2-k_s^2)\mathrm{e}^{pz}-2k^2p\mathrm{e}^{qz}}{R(k)} \mathrm{e}^{\mathrm{i}kx}\,\mathrm{d}k.
	\end{equation}
	Additionally, by deriving (\ref{equ:G_zw}) with respect to the $x$ coordinate,  we can obtain the Green's function related to the slope of the half-space: 
	\begin{equation} \label{equ:G_ztheta}
	G_{\sigma \theta}(x,z)=\frac{\partial G_{\sigma w}(x,z)}{\partial x}=\frac{\mathrm{i}}{\pi \mu} \int_{-\infty}^{\infty}\sin(kl_s/2)\frac{p(2k^2-k_s^2)\mathrm{e}^{pz}-2k^2p\mathrm{e}^{qz}}{R(k)} \mathrm{e}^{\mathrm{i}kx}\,\mathrm{d}k.
	\end{equation}
\end{subequations}

\subsubsection{Uniform shear stress distribution}

Following the same approach, we here deduce the Green's functions for a uniform shear stress distribution on the surface, as the one shown in Figure \ref{fig:fig1}c. The BCs on the free surface ($z=0$) are: 

\begin{equation} \label{equ:uniform shear stress BC at z=0}    
\sigma_{zz} (x, 0) = 0, \quad
\tau_{zx} (x, 0) = \left\{
\begin{array}{rl}
Q_x^{(n)} & \text{if}\, |x-x_n| \le l_n/2\\
0 & \text{elsewhere}
\end{array}, \right.
\end{equation}
\noindent where $Q_x^{(n)}$ is the magnitude of $\tau_{zx}$. Fourier transforming (\ref{equ:uniform shear stress BC at z=0}) yields:

\begin{equation} \label{equ:FT uniform shear stress BC at z=0}    
\bar{\sigma}_{zz} (k, 0) = 0, \quad
\bar{\tau}_{zx} (k, 0) = \frac{2Q_x^{(n)}}{k}\sin(kl_n/2)\mathrm{e}^{-\mathrm{i}kx_n}.
\end{equation}

Substituting (\ref{equ:FT uniform shear stress BC at z=0}) into (\ref{equ:general solution of u}) and (\ref{equ:general solution of w}) leads to the displacement components induced by unitary shear stress ($Q_x^{(n)}=1$ Pa) applied at the free surface, as:

\begin{subequations}
	\begin{equation} \label{equ:G_xu}
	G_{\tau u}^{(n)}(x,z)=\frac{-1}{\pi \mu} \int_{-\infty}^{\infty}\frac{\sin(kl_n/2)}{k} \frac{2k^2q\mathrm{e}^{pz}-q(2k^2-k_s^2)\mathrm{e}^{qz}}{R(k)} \mathrm{e}^{\mathrm{i}k(x-x_n)}\,\mathrm{d}k, 
	\end{equation}
	\begin{equation} \label{equ:G_xw}
	G_{\tau w}^{(n)}(x,z)=\frac{\mathrm{i}}{\pi \mu} \int_{-\infty}^{\infty}\sin(kl_n/2)  \frac{2pq\mathrm{e}^{pz}-(2k^2-k_s^2)\mathrm{e}^{qz}}{R(k)} \mathrm{e}^{\mathrm{i}k(x-x_n)}\,\mathrm{d}k.
	\end{equation}
	By deriving (\ref{equ:G_xw}) with respect to  the $x$ coordinate, we obtain the Green's function related to the slope of the half-space:
	\begin{equation} \label{equ:G_xtheta}
	G_{\tau \theta}^{(n)}(x,z)=\frac{\partial G_{\tau w}^{(n)}(x,z)}{\partial x}=\frac{-1}{\pi \mu} \int_{-\infty}^{\infty}k\sin(kl_n/2)\frac{2pq\mathrm{e}^{pz}-(2k^2-k_s^2)\mathrm{e}^{qz}}{R(k)} \mathrm{e}^{\mathrm{i}k(x-x_n)}\,\mathrm{d}k.
	\end{equation}
\end{subequations}

\subsubsection{Butterfly-shaped normal stress distribution}

At last, we derive the Green's functions for a butterfly-shaped distribution of normal stress on the surface, as shown in Figure \ref{fig:fig1}d. In this case, the BCs on the free surface ($z=0$) are:

\begin{equation} \label{equ:triangular normal stress BC at z=0}    
\sigma_{zz} (x, 0) = \left\{
\begin{array}{rl}
2\sigma_0^{(n)} (x-x_n)/l_n & \text{if}\, |x-x_n| \le l_n/2\\
0 & \text{elsewhere}
\end{array}, \right. \quad
\tau_{zx} (x, 0) = 0,
\end{equation}
\noindent in which $\sigma_0^{(n)}$ is the maximum values of $\sigma_{zz}$. Fourier transforming (\ref{equ:triangular normal stress BC at z=0}) yields:

\begin{equation} \label{equ:FT triangular normal stress BC at z=0}    
\bar{\sigma}_{zz} (k, 0) = \frac{4\mathrm{i}\sigma_0^{(n)}}{l_n k^2}\left[\frac{kl_n}{2}\cos(\frac{kl_n}{2})-\sin(\frac{kl_n}{2})\right]\mathrm{e}^{-\mathrm{i}kx_n}, \quad
\bar{\tau}_{zx} (k, 0) = 0.
\end{equation}
By substituting (\ref{equ:FT triangular normal stress BC at z=0}) into (\ref{equ:general solution of u}) and (\ref{equ:general solution of w}) we obtain the displacement components induced by a butterfly-shaped normal stress ($\sigma_0^{(n)}=1$ Pa) acting on the free surface:

\begin{subequations}
	\begin{equation} \label{equ:G_sigma0u}
	G_{\sigma_0u}^{(n)}(x,z)=\frac{-2}{\pi \mu l_n} \int_{-\infty}^{\infty}\left[\frac{kl_n}{2}\cos(\frac{kl_n}{2})-\sin(\frac{kl_n}{2})\right] \frac{(2k^2-k_s^2)\mathrm{e}^{pz}-2pq\mathrm{e}^{qz}}{kR(k)} \mathrm{e}^{\mathrm{i}k(x-x_n)}\,\mathrm{d}k, 
	\end{equation}
	\begin{equation} \label{equ:G_sigma0w}
	G_{\sigma_0w}^{(n)}(x,z)=\frac{2\mathrm{i}}{\pi \mu l_n} \int_{-\infty}^{\infty}\left[\frac{kl_n}{2}\cos(\frac{kl_n}{2})-\sin(\frac{kl_n}{2})\right] \frac{p(2k^2-k_s^2)\mathrm{e}^{pz}-2k^2p\mathrm{e}^{qz}}{k^2R(k)} \mathrm{e}^{\mathrm{i}k(x-x_n)}\,\mathrm{d}k.
	\end{equation}
	Finally, by deriving  (\ref{equ:G_sigma0w}) with respect to the $x$ coordinate we obtain:
	\begin{equation} \label{equ:G_sigma0theta}
	G_{\sigma_0\theta}^{(n)}(x,z)=\frac{\partial G_{\sigma_0w}^{(n)}(x,z)}{\partial x}=\frac{-2}{\pi \mu l_n} \int_{-\infty}^{\infty}\left[\frac{kl_n}{2}\cos(\frac{kl_n}{2})-\sin(\frac{kl_n}{2})\right] \frac{p(2k^2-k_s^2)\mathrm{e}^{pz}-2k^2p\mathrm{e}^{qz}}{kR(k)} \mathrm{e}^{\mathrm{i}k(x-x_n)}\,\mathrm{d}k.
	\end{equation}
\end{subequations}

\section{Multiple scattering formulation}\label{Multiple scattering formulation}

In this section we develop a multiple scattering formulation to quantitatively model the destructive or constructive interference between the waves generated by a harmonic distributed source, acting at the surface, and those actuated by the flexural motion of the $N$ plates. Our scope is to calculate the amplitude of the tangential and normal stresses at the base of each plate. Once these stresses are known, the total wavefield can be obtained by the superposition of the incident and scattered waves computed using the Green's functions introduced in Section \ref{Theoretical model}b.

For convenience, we define a set $\mathcal{O}=\{\boldsymbol{x}: x=x_m, z =0, m=1,...,N\} \subset \mathbb{R}^2$ to collect the plate locations. When the incident field impinges on the array of $N$ plates, the total wavefield can be expressed as the summation of incident and scattered fields:

\begin{subequations} 
	\begin{equation} \label{equ:expression of total u}
	u(x,z)=u_0(x,z)+\sum_{n=1}^{N} Q_x^{(n)} G_{\tau u}^{(n)}(x,z)+\sum_{n=1}^{N} \sigma_0^{(n)} G_{\sigma_0u}^{(n)}(x,z),
	\end{equation}
	\begin{equation} \label{equ:expression of total w}
	w(x,z)=w_0(x,z)+\sum_{n=1}^{N} Q_x^{(n)} G_{\tau w}^{(n)}(x,z)+\sum_{n=1}^{N} \sigma_0^{(n)} G_{\sigma_0w}^{(n)}(x,z),
	\end{equation}
	\begin{equation} \label{equ:expression of total rotation}
	\theta(x,z)=\theta_0(x,z)+\sum_{n=1}^{N} Q_x^{(n)} G_{\tau \theta}^{(n)}(x,z)+\sum_{n=1}^{N} \sigma_0^{(n)} G_{\sigma_0\theta}^{(n)}(x,z),
	\end{equation}
\end{subequations}
\noindent where $u_0$, $w_0$ and $\theta_0$ denote, respectively, the displacement components and rotation in the half-space due to the incident wavefield as follows:
\begin{subequations}
	\begin{equation}
	\label{eq:u0}
	u_0(x,z) = Q_0 G_{\sigma u}(x,z),
	\end{equation}
	\begin{equation}
	\label{eq:w0}
	w_0(x,z)= Q_0 G_{\sigma w}(x,z), 
	\end{equation}
	\begin{equation}
	\label{eq:theta0}
	\theta_0(x,z) = Q_0 G_{\sigma \theta}(x,z).
	\end{equation}
\end{subequations}

We substitute equations (\ref{eq:u0}, \ref{eq:theta0}) into (\ref{equ:expression of total u}) and (\ref{equ:expression of total rotation}) and specify them at the plate location $x_m$:

\begin{subequations} 
	\begin{equation} \label{equ:total u at xm}
	u(x_m,0)=Q_0 G_{\sigma u}(x_m,0)+\sum_{n=1}^{N} Q_x^{(n)} G_{\tau u}^{(n)}(x_m,0)+\sum_{n=1}^{N} \sigma_0^{(n)} G_{\sigma_0u}^{(n)}(x_m,0),
	\end{equation}
	\begin{equation} \label{equ:total rotation at xm}
	\theta(x_m,0)=Q_0 G_{\sigma \theta}(x_m,0)+\sum_{n=1}^{N} Q_x^{(n)} G_{\tau \theta}^{(n)}(x_m,0)+\sum_{n=1}^{N} \sigma_0^{(n)} G_{\sigma_0\theta}^{(n)}(x_m,0).
	\end{equation}
\end{subequations}

Similarly, we consider the shear and normal stresses at the base of the plate, i.e., (\ref{equ:shear stress from bending moment}) and (\ref{equ:normal stress from bending moment}), respectively, and express them in force of (\ref{equ:uniform shear stress BC at z=0}) and (\ref{equ:triangular normal stress BC at z=0}):
\begin{subequations} 
	\begin{equation} \label{equ:Qx at xm}
	Q_x^{(m)}=\Omega_1^{(m)}u(x_m,0)-\Omega_2^{(m)}\theta(x_m,0),
	\end{equation}
	\begin{equation} \label{equ:sigma_0 at xm}
	2\sigma_0^{(m)}/l_m = \Omega_3^{(m)}u(x_m,0)-\Omega_4^{(m)}\theta(x_m,0).
	\end{equation}
\end{subequations}
At this stage, we can express the displacements $u(x_m,0)$ and $\theta(x_m,0)$ in (\ref{equ:Qx at xm}) and (\ref{equ:sigma_0 at xm}) by means of (\ref{equ:total u at xm}) and (\ref{equ:total rotation at xm}) to obtain:
\begin{subequations} 
	\begin{equation} \label{equ:Qx at xm 2}
	\begin{split}
	Q_x^{(m)} &= Q_0\left[\Omega_1^{(m)}G_{\sigma u}(x_m,0)-\Omega_2^{(m)}G_{\sigma \theta}(x_m,0)\right] \\ 
	&+ \sum_{n=1}^{N} Q_x^{(n)}\left[\Omega_1^{(m)}G_{\tau u}^{(n)}(x_m,0)-\Omega_2^{(m)}G_{\tau \theta}^{(n)}(x_m,0)\right] \\
	&+\sum_{n=1}^{N} \sigma_0^{(n)}\left[\Omega_1^{(m)}G_{\sigma_0 u}^{(n)}(x_m,0)-\Omega_2^{(m)}G_{\sigma_0 \theta}^{(n)}(x_m,0)\right],
	\end{split}
	\end{equation}
	\begin{equation} \label{equ:sigma_0 at xm 2}
	\begin{split}
	2\sigma_0^{(m)}/l_m &= Q_0\left[\Omega_3^{(m)}G_{\sigma u}(x_m,0)-\Omega_4^{(m)}G_{\sigma \theta}(x_m,0)\right] \\ 
	&+ \sum_{n=1}^{N} Q_x^{(n)}\left[\Omega_3^{(m)}G_{\tau u}^{(n)}(x_m,0)-\Omega_4^{(m)}G_{\tau \theta}^{(n)}(x_m,0)\right] \\
	&+\sum_{n=1}^{N} \sigma_0^{(n)}\left[\Omega_3^{(m)}G_{\sigma_0 u}^{(n)}(x_m,0)-\Omega_4^{(m)}G_{\sigma_0 \theta}^{(n)}(x_m,0)\right].
	\end{split}
	\end{equation}
\end{subequations}
With some algebra, equations (\ref{equ:Qx at xm 2}) and (\ref{equ:sigma_0 at xm 2}) can be reorganized as:

\begin{subequations}
	\begin{equation} \label{equ:shear stress at xm}
	\begin{split}
	-Q_0\left[\Omega_1^{(m)}G_{\sigma u}(x_m,0)-\Omega_2^{(m)}G_{\sigma \theta}(x_m,0)\right] &= \sum_{n=1}^{N} Q_x^{(n)}\left[\Omega_1^{(m)}G_{\tau u}^{(n)}(x_m,0)-\Omega_2^{(m)}G_{\tau \theta}^{(n)}(x_m,0)-\delta_{n}^{m}\right] \\
	&+\sum_{n=1}^{N} \sigma_0^{(n)}\left[\Omega_1^{(m)}G_{\sigma_0 u}^{(n)}(x_m,0)-\Omega_2^{(m)}G_{\sigma_0 \theta}^{(n)}(x_m,0)\right],
	\end{split}
	\end{equation}
	\begin{equation} \label{equ:triangular normal stress at xm}
	\begin{split}
	-Q_0\left[\Omega_3^{(m)}G_{\sigma u}(x_m,0)-\Omega_4^{(m)}G_{\sigma \theta}(x_m,0)\right] &= \sum_{n=1}^{N} Q_x^{(n)}\left[\Omega_3^{(m)}G_{\tau u}^{(n)}(x_m,0)-\Omega_4^{(m)}G_{\tau \theta}^{(n)}(x_m,0)\right] \\
	&+\sum_{n=1}^{N} \sigma_0^{(n)}\left[\Omega_3^{(m)}G_{\sigma_0 u}^{(n)}(x_m,0)-\Omega_4^{(m)}G_{\sigma_0 \theta}^{(n)}(x_m,0)-2\delta_{n}^{m}/l_n\right].
	\end{split}
	\end{equation}
\end{subequations}

\noindent where $\delta_{n}^{m}$ is the Kronecker delta.

Equations (\ref{equ:shear stress at xm}) and (\ref{equ:triangular normal stress at xm}), applied to $\boldsymbol{x} \in \mathcal{O}$, provide a system of $2\times N$ non homogeneous linear equations, in the  $2\times N$ unknown coefficients $Q_x^{(n)}$ and $\sigma_0^{(n)}$, written in compact form as:

\begin{equation}
\boldsymbol{S} \boldsymbol{y}=\boldsymbol{b},
\label{equ:system compact form}
\end{equation}
\noindent in which the corresponding coefficients are:

\begin{equation}
\boldsymbol{S} = \left[\begin{array}{cc}
{\boldsymbol{S}_{11}} & {\boldsymbol{S}_{12}}  \\
{\boldsymbol{S}_{21}} & {\boldsymbol{S}_{22}}
\end{array}\right], \quad 
\boldsymbol{y} = \left[\begin{array}{c}
{Q_x^{(n)}} \\
{\sigma_0^{(n)}} 
\end{array}\right], \quad 
\boldsymbol{b} = \left[\begin{array}{c}
{\boldsymbol{b}_1} \\
{\boldsymbol{b}_2} 
\end{array}\right],
\end{equation}
\noindent with components:
\begin{subequations}
	\begin{equation}
	\boldsymbol{S}_{11}=\sum_{n=1}^{N} \left[\Omega_1^{(m)}G_{\tau u}^{(n)}(x_m,0)-\Omega_2^{(m)}G_{\tau \theta}^{(n)}(x_m,0)-\delta_{n}^{m}\right],
	\end{equation}
	
	\begin{equation}
	\boldsymbol{S}_{12}=\sum_{n=1}^{N} \left[\Omega_1^{(m)}G_{\sigma_0 u}^{(n)}(x_m,0)-\Omega_2^{(m)}G_{\sigma_0 \theta}^{(n)}(x_m,0)\right],
	\end{equation}
	
	\begin{equation}
	\boldsymbol{S}_{21}=\sum_{n=1}^{N} \left[\Omega_3^{(m)}G_{\tau u}^{(n)}(x_m,0)-\Omega_4^{(m)}G_{\tau \theta}^{(n)}(x_m,0)\right],
	\end{equation}
	
	\begin{equation}
	\boldsymbol{S}_{22}=\sum_{n=1}^{N} \left[\Omega_3^{(m)}G_{\sigma_0 u}^{(n)}(x_m,0)-\Omega_4^{(m)}G_{\sigma_0 \theta}^{(n)}(x_m,0)-2\delta_{n}^{m}/l_n\right],
	\end{equation}
	
	\begin{equation}
	\boldsymbol{b}_1=-Q_0\left[\Omega_1^{(m)}G_{\sigma u}(x_m,0)-\Omega_2^{(m)}G_{\sigma \theta}(x_m,0)\right],
	\end{equation}
	
	\begin{equation}
	\boldsymbol{b}_2=-Q_0\left[\Omega_3^{(m)}G_{\sigma u}(x_m,0)-\Omega_4^{(m)}G_{\sigma \theta}(x_m,0)\right].
	\end{equation}
	
\end{subequations}
To avoid numerical instabilities in the solution of \eqref{equ:system compact form}, we assume a small hysteretic damping ratio $\zeta = 0.1\%$ in both the substrate and the plates elastic response to remove the poles of the integrands in the Green's functions. 

The computed coefficients $Q_x^{(n)}$ and $\sigma_0^{(n)}$ are finally used into (\ref{equ:expression of total u}) and (\ref{equ:expression of total w}) to obtain the total wavefield in the half-space.

\section{Numerical examples} \label{Numerical examples}

In this section, we assess the validity of the proposed multiple scattering formulation, namely (\ref{equ:expression of total u}, \ref{equ:expression of total w}) by modeling three case studies. The first case considers the interaction of  Rayleigh waves with a single plate; in the second example, we model the interaction between five distinct plates arranged in a random configuration; the last scenario describes the response of a finite-size metasurface composed by 30 plates. The mechanical and geometrical parameters of the three examples are collected in Table \ref{tab:tab1}.

\begin{table}[htbp]
	\caption{Mechanical parameters for plates and the elastic half-space \cite{wootton2019asymptotic}.}
	\label{tab:tab1}
	\begin{tabular*}{\hsize}{@{}@{\extracolsep{\fill}}lll@{}}
		\toprule
		Symbol & Definition & Value  \\   
		\midrule
		$Q_0$ & Distributed source amplitude   & 1 MPa \\
		$l_s$ & Footprint thickness of the source  & 1 m  \\
		$\rho$ & Mass density of half-space  & 13000 kg/m$^3$ \\
		$\lambda$ & First Lam\'e constant of half-space & 702 MPa \\
		$\mu$ & Shear modulus of half-space & 325 MPa \\
		$l_n$ & Plate thickness  & 0.3 m  \\
		$\rho_n$ & Plate density  & 450 kg/m$^3$ \\
		$E_n$ & Young modulus of plate & 1547 MPa \\
		$\nu_n$ & Poisson ratio of plate & 0.3 \\
		$\zeta$ & Hysteretic damping ratio & 0.1$\%$ \\
		\bottomrule
	\end{tabular*}
\end{table}

\begin{figure}[htbp]
	\centering
	\includegraphics[width=5 in]{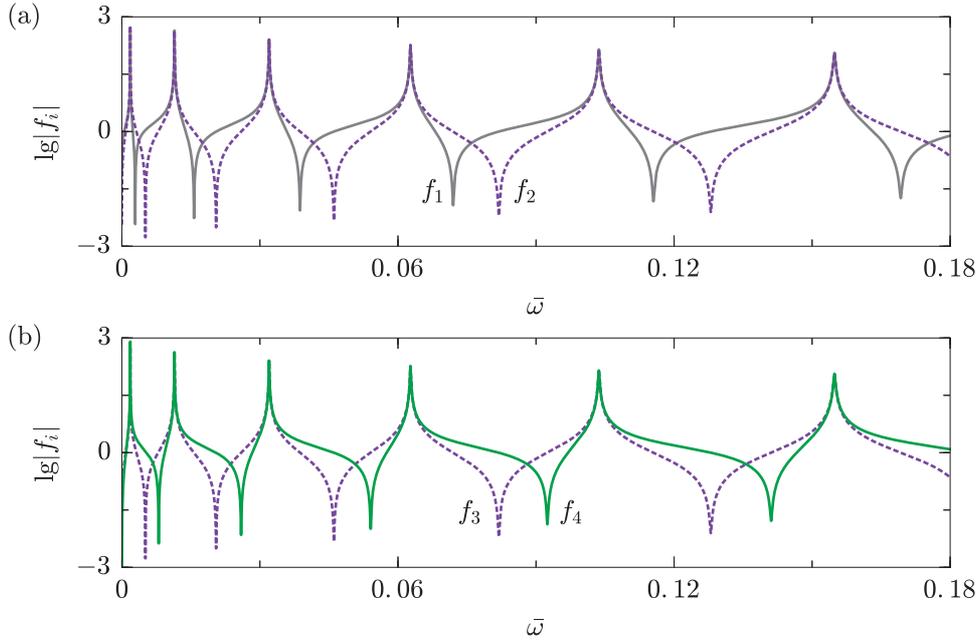}
	\caption{Variations of the impedance parameters vs. frequency ($l = 0.3\; \mathrm{m}$ and $h = 14\; \mathrm{m}$).}
	\label{fig:fig2}
\end{figure}

Before delving into the description of the case studies, let us discuss the impedance properties derived in (\ref{equ:Omega_1}-\ref{equ:Omega_4}) which depend on the dimensionless functions $f_i \;(i=1,...,4)$. The variations of these dimensionless functions with respect to the frequency are reported in Figure \ref{fig:fig2}, where the dimensionless frequency $\bar{\omega}=\omega l/(\pi c_s)$ is used \cite{achenbach1973wave}. According to  (\ref{equ:shear stress from bending moment}), the shear stress provided by the plate flexural motion is determined by both $f_1$ and $f_2$, here collected in Figure \ref{fig:fig2}a. In the frequency range $\bar{\omega}=[0,0.18]$ the impedance parameters  $f_1$ and $f_2$ present six identical resonance peaks, which correspond to the first six flexural resonances of a cantilever plate, i.e., $1+\cosh(\beta h)\cos(\beta h)=0$. Strong interaction effects between the plate and the half-space dynamics are expected in the vicinity of these flexural resonances, while minimal coupling occurs at the valley points. It should be noted that the minimal values of $f_1$ and $f_2$ do not overlap except for the static condition (zero frequency), suggesting that the flexural motion of the plate  always exchange non-zero shear stresses with the surface half-space. The same arguments apply to the dimensionless impedance functions $f_3$ and $f_4$, reported in Figure \ref{fig:fig2}b, which govern the normal stress exerted by the plate on the half-space.

\subsection{Single plate scenario} \label{Single plate scenario}

\begin{figure}[htbp]
	\centering
	\includegraphics[width=5 in]{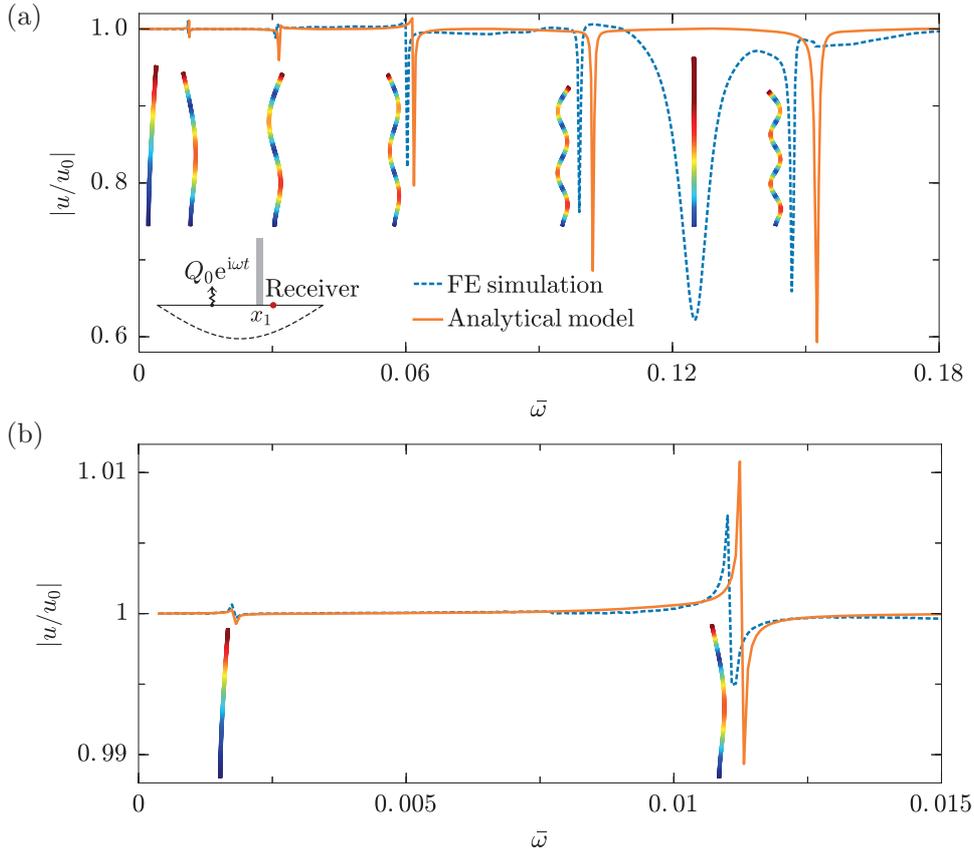}
	\caption{(a) Amplitude ratio $A_R=|u/u_0|$ of a single plate ($l_1 = 0.3\; \mathrm{m}$ and $h_1 = 14\; \mathrm{m}$). The plate is located at $x_1=100\; \mathrm{m}$ and the receiver $x=120\; \mathrm{m}$. (b) Zoomed-in amplitude ratio in the low-frequency regime.} 
	\label{fig:fig3}
\end{figure}

A thin plate with $l_1 = 0.3\; \mathrm{m}$ and $h_1 = 14\; \mathrm{m}$ is located at $x_1 = 100\; \mathrm{m}, z = 0\; \mathrm{m}$ and is excited by a harmonic distributed source with footprint thickness $l_s = 1\; \mathrm{m}$ and amplitude $Q_0 = 1\; \mathrm{MPa}$ located at the origin of the coordinate system. Figure \ref{fig:fig3}a shows the amplitude ratio $A_R=|u/u_0|$ of the horizontal displacement at $x = 120\; \mathrm{m}, z = 0\; \mathrm{m}$ for $\bar{\omega} \in [0,0.18]$ which includes the first six flexural resonances.

For comparison, we provide the same amplitude ratio computed via a 2D FE simulation (dashed line). In the FE environment, both the plate and the half-space are modeled using 2D plane strain elasticity (see \ref{Appendix A} for details on the model). Nonetheless, a very good agreement between the results obtained by using the proposed formulation and the FE is found. One can observe that the interaction between the plate and the half-space in the vicinity of the plate resonances yields minimal values of the amplitude ratio. The drops in the low-frequency range can be better appreciated in Figure \ref{fig:fig3}b, where the amplitude ratio is shown in the frequency range $\bar{\omega} \in [0,0.015]$.

Note that the flexural resonances computed by the analytical model occur at frequency values slightly higher than the actual 2D FE results. Indeed, our formulation based on the  Kirchhoff plate theory neglects both the shear deformation and rotary inertia, making the model more rigid than the 2D plane strain model. Besides, we remind that the proposed formulation does not capture the axial resonance of the plate, i.e. the valley between the fifth and sixth bending resonance, since we consider the flexural motion of the plate only. The reader can appreciate the nature of the different resonant modes from the modal shapes provided in the figure inset.

Additionally, let us remark that the proposed formulation reduces significantly the computational cost needed to obtain the amplitude ratio, as it can compute the output at a desired receiver location  without the need for a discretization of the whole model, as in the case of FE models. For example, with a personal computer, it takes only a few seconds to simulate the system in Figure \ref{fig:fig3} while around one hour for the FE model.

\subsection{Randomly distributed plates}
\label{Randomly distributed plates}

\begin{figure}[htbp]
	\centering
	\includegraphics[width=5 in]{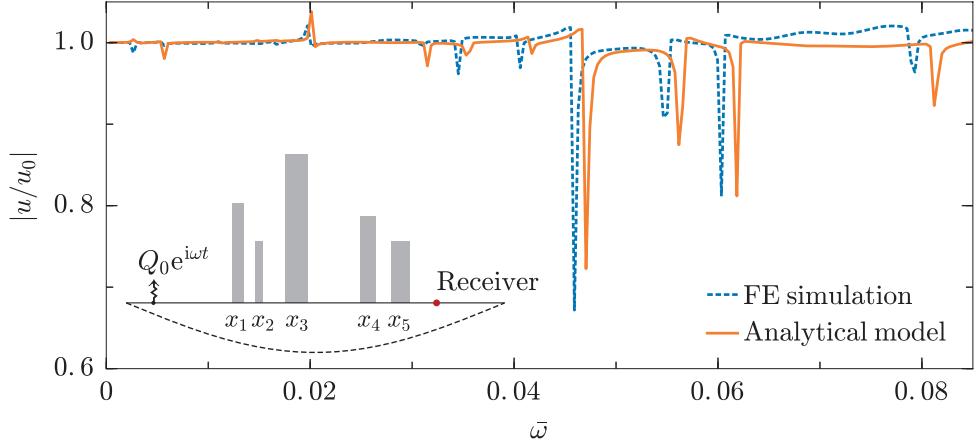}
	\caption{Amplitude ratio $A_R=|u/u_0|$ for a configuration of five randomly distributed plates. The receiver is located at $x=130\; \mathrm{m}$.}
	\label{fig:fig4}
\end{figure}

As a second example, we consider a system of five different plates arranged atop the half-space in the arbitrary configuration shown in the inset of Figure \ref{fig:fig4}. Random arrangements of flexural resonators are often found in urban vibration problems, i.e., where a series of buildings are impacted by seismic waves and their responses are affected by those of the other buildings \cite{alexander2013simple}.  We show that an arbitrary configuration of flexural resonators can be readily modeled by our formulation.

The five plates are characterized by the following geometrical parameters: $l=[0.3, 0.2, 0.6, 0.4, 0.5]\; \mathrm{m}$, $h=[14, 10, 16, 12, 10]\; \mathrm{m}$, and location $x=[100, 102, 106, 116, 118]\; \mathrm{m}$. The amplitude ratio  at $x = 130\; \mathrm{m}, z = 0\; \mathrm{m}$ for $\bar{\omega} \in [0,0.085]$, as calculated by our formulation, is shown in Figure \ref{fig:fig4} (solid line). As for the single plate scenario, the interaction between the plates resonances and the substrate wavefield results in multiple drops of the amplitude ratio, located at the plate resonances. For comparison, we also provide the results of a FE simulation (dashed line), developed using 2D plane-strain elasticity model, which well agree with those of the analytical framework. 

\begin{figure}[htbp]
	\centering
	\includegraphics[width=5 in]{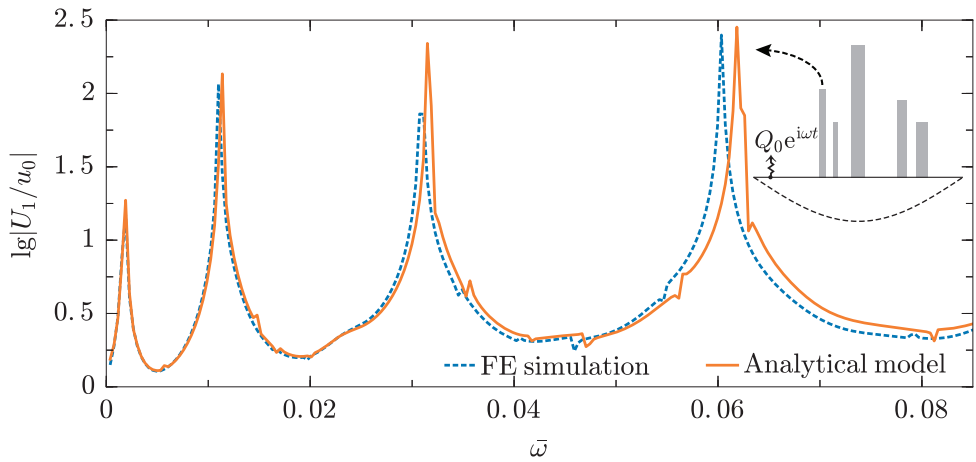}
	\caption{Horizontal response spectrum at the free end of the first plate $U_1(h_1)$.}
	\label{fig:fig5}
\end{figure}

Finally, in Figure \ref{fig:fig5} we show the horizontal response of the top of the first plate, i.e., $U_1(h_1)$, calculated using our formulation (solid line). In particular, the response at the top of the plate is found via equation (\ref{equ:general solution of flexural motion}) once the base displacement and rotation, namely $u(x_1,0)$ and $\theta(x_1,0)$, are computed. It can be observed that the four peaks in the spectrum match perfectly with the first four resonances of the plate (see Figure \ref{fig:fig3}). Additionally, we still observe some small valleys in the spectrum, which evidence the mutual interactions among these plates \cite{PU2021103547}. The analytical solutions are again validated with FE simulations (dashed line).

\subsection{Finite-size metasurfaces}

We now consider the propagation of Rayleigh waves across a cluster of plates, i.e., a  flexural metasurface. We aim to show the capability of the proposed formulation to reproduce the wavefield in this complex configuration. 

Thus, we consider an array of 30 identical plates ($l = 0.3\; \mathrm{m}$ and $h = 14\; \mathrm{m}$), which are arranged periodically with a lattice constant $L=2\; \mathrm{m}$. A similar configuration was considered in Ref. \cite{colquitt2017seismic} for Rayleigh waves interacting with rods. To predict the effect of flexural resonators on the propagation of Rayleigh waves, we first compute the dispersion relation for the infinite system. The dispersion relation can be obtained either analytically, by exploiting the thin-plates impedances, or via 2D FE models imposing periodic BCs, as shown in Figure \ref{fig:fig6}a (more details on the computation can be found in \ref{Appendix B}). 

In particular, we show with solid black lines the solution obtained from the analytical dispersion equation (\ref{equ:dispersion equation}) and with orange circles the dispersion relation calculated by FE. The dashed blue line indicates the shear wave sound-line. Overall, a good agreement is found between the analytical and FE computed dispersion curves. Note that, differently from a metasurface of longitudinal resonators \cite{colquitt2017seismic}, the bandgap induced by the plate flexural resonance is barely visible in this scenario. Indeed, the bandgap width is related to the slenderness of the plate as well as the stiffness ratio of the plate to the substrate. For further details, the reader can refer to \cite{colquitt2017seismic, marigo2020surface}. Nevertheless, for incident waves at one of the plate resonance, we still expect a strong localization of the wavefield at the onset of the metasurface, as a result of the interaction between the incident field and the plate flexural motion.

To visualize this phenomenon, we consider an incident Rayleigh wave at the fourth flexural resonance of the plate ($\bar{\omega}=0.06$), which is excited by the source $Q_0 \mathrm{e}^{\mathrm{i}\omega t}$ placed 100 m away from the first plate. The wavefield, calculated by (\ref{equ:expression of total u}) and (\ref{equ:expression of total w}), is displayed in Figure \ref{fig:fig6}b and highlights the strong interaction between the Rayleigh waves and the plates. As shown in Figure \ref{fig:fig6}c, the predicted wavefield is in very good agreement with the one computed by using the FE simulation.

\begin{figure}[htbp]
	\centering
	\includegraphics[width=5.5 in]{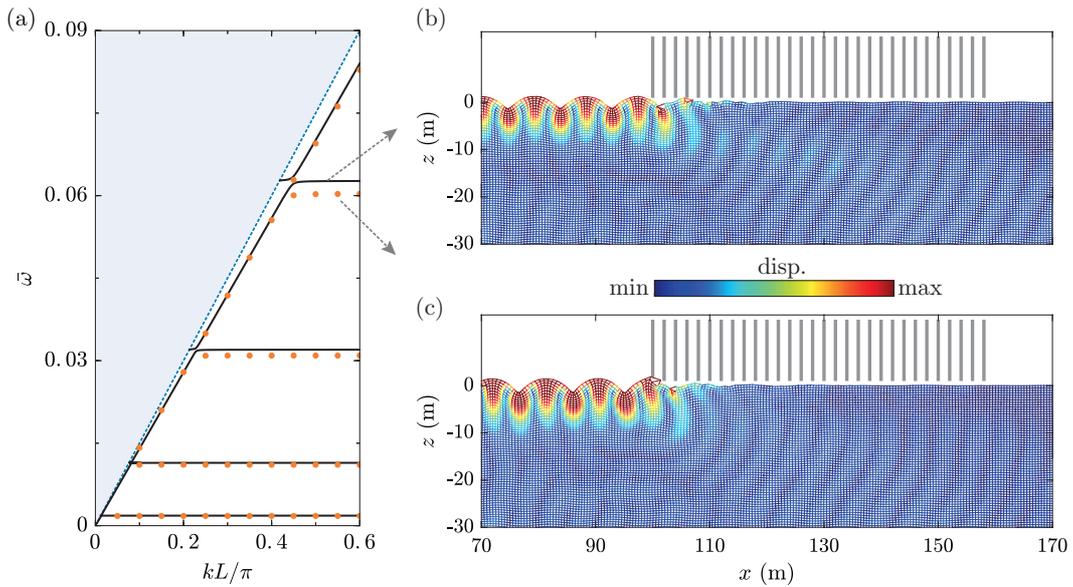}
	\caption{Rayleigh wave interacting with a metasurface atop an elastic half-space. (a) The dispersion curves for an infinite array of plates resting on the half-space ($L=2\; \mathrm{m}$). Solid lines and circles denote, respectively, the analytical solution and the FE simulation, while the dashed line indicates the shear wave sound-line. (b) Analytical solution and (c) FE simulation of the wavefield at the fourth flexural resonance for the half-space supporting a periodic array of $N=30$ plates.}
	\label{fig:fig6}
\end{figure}

\section{Conclusion} \label{Conclusion}

We have provided a multiple scattering formulation to investigate the interaction of Rayleigh waves with an array of thin plates located at the surface of an elastic half-space. The presence of the plates on top of the half-space is modeled by means of their equivalent impedances, derived by exploiting the Kirchhoff plate theory. The flexural motion of the plates, under the action of Rayleigh waves, provides additional scattered wavefields which are modeled invoking ad-hoc Green's functions. The amplitudes of the scattered wavefields are obtained from a multiple scattering formulation by imposing the continuity of displacement and slope at the footprint of the thin plates. Our approach enables the treatment of a flexural metasurface in an arbitrary way, namely with no restriction on the mechanical and geometrical properties of the plates as well as in their spatial configuration. The capability of the method has been discussed by investigating a single and an array of thin plates atop a homogeneous half-space. The methodology can capture the interaction between the plate dynamics, mediated by the elastic substrate, and the complex wave patterns induced by the presence of multiple flexural resonators. Further generalization of this approach could consider the effect of the axial motion in the plate dynamics and its interplay with the bending motion in the high-frequency regime.   

\section*{CRediT authorship contribution statement}
\noindent \textbf{Xingbo Pu:} Conceptualization, Methodology, Investigation, Software, Writing - original draft. \textbf{Antonio Palermo:} Conceptualization, Investigation, Validation, Writing - review \& editing, Co-supervision. \textbf{Alessandro Marzani:} Conceptualization, Investigation, Writing - review \& editing, Supervision, Funding acquisition.

\section*{Declaration of competing interest}
\noindent The authors declare that they have no conflict of interest.

\section*{Acknowledgments}
\noindent This project has received funding from the European Union’s Horizon 2020 research and innovation programme under the Marie Skłodowska Curie grant agreement No 813424.

\appendix

\section{Details on the 2D FE model} \label{Appendix A}

In this Appendix, we provide the details of the 2D plane-strain FE model, built in COMSOL Multiphysics, used to verify the amplitude ratio $A_R$ obtained by using our analytical solution in Section \ref{Single plate scenario}. The FE model comprises  a plate and a substrate with in-plane dimensions $H\times W$ (see Figure \ref{fig:figA1}). For low-frequency  harmonic simulations $\bar{\omega} \in [0,0.015]$, we set  $H=300\; \mathrm{m}$ and $W=500\; \mathrm{m}$ to properly model the long-wavelength elastic waves generated by the source.
For  higher frequency harmonic simulations $\bar{\omega} \in [0.015,0.18]$, we set  $H=80\; \mathrm{m}$ and $W=240\; \mathrm{m}$ to reduce the computational burden required to model the system. Both the plate and the substrate are discretized using a fine mesh of quadratic triangular elements, which ensure convergent results at the highest frequency of interest.

Perfectly matched layers (PML) are added around the substrate to minimize reflections from the domain borders and simulate the response of a half-space.

\setcounter{figure}{0}
\begin{figure}[htbp]
	\centering
	\includegraphics[width=4.5 in]{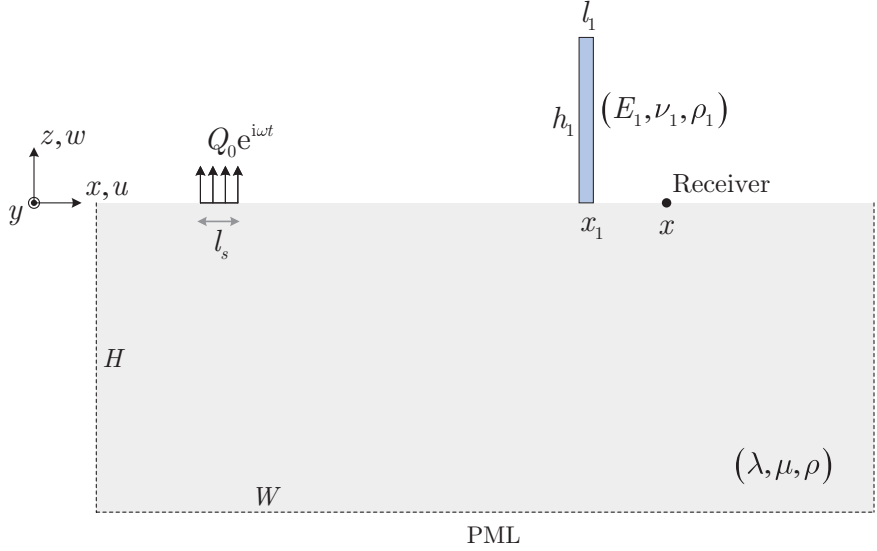}
	\caption{Schematic of the 2D FE model.}
	\label{fig:figA1}
\end{figure}

Analogous FE models are used to compute the numerical amplitude ratio for the five plates configuration discussed in Section \ref{Randomly distributed plates} and the wavefield of Figure \ref{fig:fig6}c.

\section{Analytical vs. FE dispersion relation} \label{Appendix B}

In this Appendix, we provide the dispersion relation for Rayleigh-like surface waves existing in an elastic half-space equipped with an array of periodic thin-plates. For convenience, we introduce the set $\mathcal{H}_n=\{\boldsymbol{x}: |x-x_n| \le L/2, z \le 0\} \subset \mathbb{R}^2$ to denote the half-space portion enclosed within the generic $n$-th unit cell.

\setcounter{figure}{0}
\begin{figure}[htbp]
	\centering
	\includegraphics[width=5.5 in]{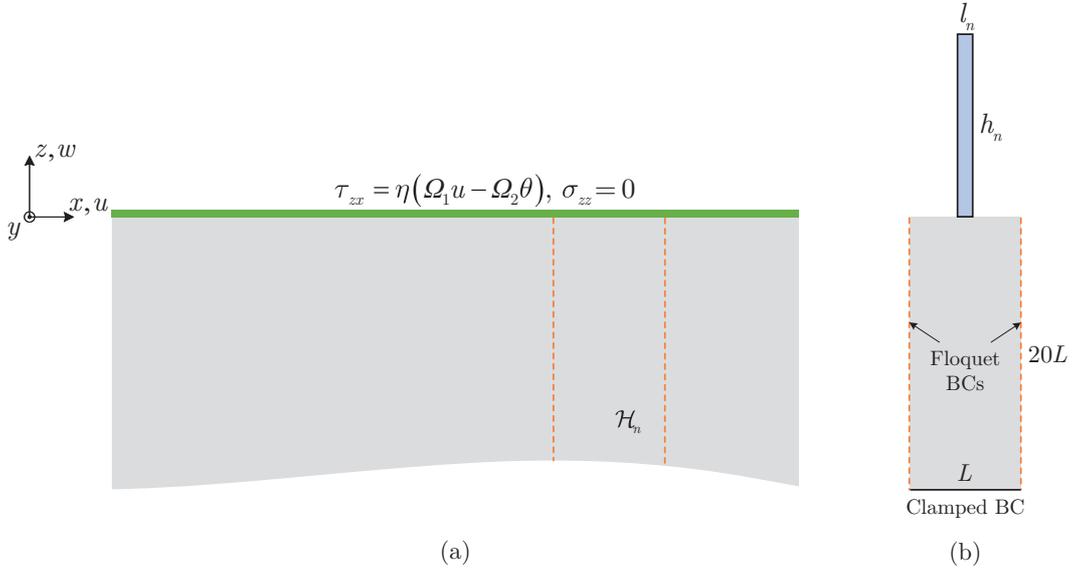}
	\caption{Schematics considered to model the dispersion relation for Rayleigh waves interacting with a periodic array of thin plates with flexural resonances: (a) analytical model and (b) FE model.}
	\label{fig:figB1}
\end{figure}

Since we are seeking for wave solutions of the Rayleigh type, the potentials $\varphi$ and $\psi_y$ can be assumed as \cite{achenbach1973wave}:

\begin{equation} \label{equ:Rayleigh wave potentials}
\varphi(\boldsymbol{x})=A \mathrm{e}^{pz+\mathrm{i} kx}, \quad \psi_y(\boldsymbol{x})=B \mathrm{e}^{qz+\mathrm{i} kx}, \quad \text{for} \quad \boldsymbol{x} \in \mathcal{H}_n.
\end{equation}
\noindent
where the wave numbers $p$ and $q$ are given in (\ref{equ:wave numbers of potentials}).

Accordingly, the displacement components and the slope of the half-space surface can be written as:
\begin{subequations} 
	\begin{equation} \label{equ:u at z=0}
	u(x,0)=(\mathrm{i}kA-qB)\mathrm{e}^{\mathrm{i} kx},
	\end{equation}
	\begin{equation} \label{equ:w at z=0}
	w(x,0)=(pA+\mathrm{i}kB)\mathrm{e}^{\mathrm{i} kx},
	\end{equation}
	\begin{equation} \label{equ:rotation at z=0}
	\theta(x,0)=\frac{\partial w(x,0)}{\partial x}=(\mathrm{i} kp A-k^2 B)\mathrm{e}^{\mathrm{i} kx},
	\end{equation}
\end{subequations}
Similarly, the stress components at the half-space surface read:
\begin{subequations}
	\begin{equation} \label{equ:shear stress at z=0}
	\tau_{zx}(x,0)=\mu[2 \mathrm{i} k p A+(k_{s}^{2}-2 k^{2}) B]\mathrm{e}^{\mathrm{i} kx},
	\end{equation}
	\begin{equation} \label{equ:normal stress at z=0}
	\sigma_{zz}(x,0)=\mu [-(k_{s}^2-2 k^2) A+2 \mathrm{i} kq B]\mathrm{e}^{\mathrm{i} kx}.
	\end{equation}
\end{subequations}
The presence of a periodic arrangement of flexural resonators is treated via an effective medium approach \cite{boechler2013interaction}. To this purpose we average the shear stress (\ref{equ:shear stress from bending moment}) and normal stress (\ref{equ:normal stress from bending moment}) exerted by the plate over the half-space, along the whole surface of the unit cell:

\begin{subequations}
	\begin{equation} \label{equ:effective shear stress at z=0}
	\tilde{\tau}_{zx}(x,0)=\frac{1}{L}\int_{x_n-l_n/2}^{x_n+l_n/2}\tau_{zx}^{(n)}(x,0)\,\mathrm{d}x= \eta [\Omega_1 u(x,0) - \Omega_2 \theta(x,0)],
	\end{equation}
	\begin{equation} \label{equ:effective normal stress at z=0}
	\tilde{\sigma}_{zz}(x,0)=\frac{1}{L}\int_{x_n-l_n/2}^{x_n+l_n/2}\sigma_{zz}^{(n)}(x,0)\,\mathrm{d}x=0.
	\end{equation}
\end{subequations}
\noindent
in which $\eta=l_n/L$ is the ratio between the cross-section of the plate to the unit cell. As per our assumptions, the flexural motion provides an average null normal stress.

At this stage, the dispersion law is obtained by imposing the stress continuity at the half-space surface i.e., $\tau_{zx}(x,0)=\tilde{\tau}_{zx}(x,0)$, $\sigma_{zz}(x,0)=\tilde{\sigma}_{zz}(x,0)$.
In particular, we substitute the displacements in (\ref{equ:u at z=0}) and (\ref{equ:rotation at z=0}) in (\ref{equ:effective shear stress at z=0})-(\ref{equ:effective normal stress at z=0}), and equate the obtained expressions with (\ref{equ:shear stress at z=0}-\ref{equ:normal stress at z=0}).
With some algebra, we  obtain the following system:

\begin{equation} \label{equ:coupled BC at z=0}
\left[\begin{array}{cc}
\mathrm{i}k(\eta \Omega_1-\eta \Omega_2 p-2\mu p) &
-\mu(k_{s}^2-2 k^2)-\eta \Omega_1 q+\eta \Omega_2 k^2 \\
\mu (k_{s}^2-2 k^2) & -2\mathrm{i} \mu kq\end{array}\right] 
\left[\begin{array}{l}
A \\
B
\end{array}\right]=\bf{0}.
\end{equation}

Nontrivial solutions of (\ref{equ:coupled BC at z=0}) yield the dispersion relation for Rayleigh waves in such a system, accounting for flexural resonances:

\begin{equation} \label{equ:dispersion equation}
2\xi^2 \sqrt{\xi^2-1} \left(2\sqrt{\xi^2-r^2}+\frac{\eta \Omega_2}{\mu}\sqrt{\xi^2-r^2}- \frac{\eta \Omega_1}{\rho c_s \omega}\right)
=\left(1-2\xi^2\right) \left( 1-2\xi^2-\frac{\eta \Omega_2}{\mu}\xi^2 +\frac{\eta \Omega_1}{\rho c_s \omega}\sqrt{\xi^2-1}\right).
\end{equation}

\noindent
where $\xi=k c_s/\omega$ and $r=c_s/c_p$. Solutions of \eqref{equ:dispersion equation} are  computed using the Newton-Raphson method. 

To validate the analytical dispersion equation (\ref{equ:dispersion equation}), we develop a 2D plane-strain FE model using 2D elasticity in COMSOL Multiphysics (Figure \ref{fig:figB1}b). Following the approach in Ref. \cite{palermo2018metabarriers} developed for mass-spring metasurfaces, we build a unit cell considering the elastic substrate with dimensions $L \times 20L$. To simulate the dynamics of an infinite array of periodic plates, we impose a pair of Floquet periodic BCs  on the vertical substrate edges. A clamped BC is enforced at the bottom edge to avoid rigid motions. The numerical dispersion curves are obtained by solving the eigenvalue problem for wave number in the Floquet BCs varying between $k=[0, \pi/L]$.

\bibliography{mybibfile}

\end{document}